\documentclass{article}

\usepackage[dblblindworkshop, final]{neurips_2025}
\workshoptitle{Reliable ML from Unreliable Data}

\usepackage[utf8]{inputenc} 
\usepackage[T1]{fontenc}    
\usepackage{hyperref}       
\usepackage{url}            
\usepackage{booktabs}       
\usepackage{amsfonts}       
\usepackage{nicefrac}       
\usepackage{microtype}      
\usepackage{xcolor}         
\usepackage{booktabs}       
\usepackage{graphicx}
\usepackage{algorithm}      
\usepackage{algpseudocode}  

\title{Quantifying CBRN Risk in Frontier Models}

\author{%
Divyanshu Kumar \\
Enkrypt AI \\
\texttt{divyanshu@enkryptai.com}
\And
Nitin Aravind Birur \\
Enkrypt AI \\
\texttt{nitin@enkryptai.com} \And
Tanay Baswa \\
Enkrypt AI \\
\texttt{tanay@enkryptai.com} \And
Sahil Agarwal \\
Enkrypt AI \\
\texttt{sahil@enkryptai.com} \And
Prashanth Harshangi \\
Enkrypt AI \\ 
\texttt{prashanth@enkryptai.com}
}

\begin{document}

\maketitle

\begin{abstract}
  Frontier Large Language Models (LLMs) pose unprecedented dual-use risks through the potential proliferation of chemical, biological, radiological, and nuclear (CBRN) weapons knowledge. We present the first comprehensive evaluation of 10 leading commercial LLMs against both a novel 200-prompt CBRN dataset and a 180-prompt subset of the FORTRESS benchmark, using a rigorous three-tier attack methodology. Our findings expose critical safety vulnerabilities: Deep Inception attacks achieve 86.0\% success versus 33.8\% for direct requests, demonstrating superficial filtering mechanisms; Model safety performance varies dramatically from 2\% (claude-opus-4) to 96\% (mistral-small-latest) attack success rates; and eight models exceed 70\% vulnerability when asked to enhance dangerous material properties. We identify fundamental brittleness in current safety alignment, where simple prompt engineering techniques bypass safeguards for dangerous CBRN information. These results challenge industry safety claims and highlight urgent needs for standardized evaluation frameworks, transparent safety metrics, and more robust alignment techniques to mitigate catastrophic misuse risks while preserving beneficial capabilities.
\end{abstract}

\section{Introduction}

The rapid advancement of Large Language Models (LLMs) presents a significant dual-use challenge within artificial intelligence research. Although these models offer substantial benefits for scientific inquiry, medical research, and educational applications, they simultaneously introduce potential risks regarding the proliferation of chemical, biological, radiological, and nuclear (CBRN) weapons knowledge. This concern has garnered attention from governmental bodies, as evidenced by the U.S. Executive Order 14110 on "Safe, Secure, and Trustworthy AI" and subsequent Department of Homeland Security analysis indicating that LLMs may reduce barriers for malicious actors seeking to develop CBRN threats \cite{dhs2024}.

Despite widespread acknowledgment of CBRN risks within AI safety frameworks, a significant research gap persists in empirical assessment of safety measure efficacy against sophisticated adversarial techniques. Current evaluation methodologies exhibit several limitations: they predominantly employ direct prompting techniques that inadequately represent realistic adversarial behaviors; they disproportionately focus on biological risks at the expense of chemical, radiological, and nuclear domains; and they primarily assess factual knowledge retrieval rather than evaluating models' capacity to facilitate applications of potentially harmful information.

This study addresses these methodological gaps through systematic evaluation of 10 frontier LLMs, employing:

\begin{enumerate}
    \item A comprehensive evaluation dataset comprising 200 prompts in all CBRN domains, designed to assess multiple capabilities dimensions including factual recall, process instruction, novel generation, and synthesis guidance.
    
    \item A structured three-tier attack taxonomy of increasing methodological sophistication involving direct requests, obfuscated requests, and Deep Inception attacks to simulate realistic adversarial approaches.
    
    \item Standardized evaluation criteria aligned with established AI risk management frameworks to allow systematic cross-model comparison and identification of vulnerability patterns.
\end{enumerate}

Our empirical analysis reveals several significant findings regarding the current implementation of safety:

\begin{enumerate}
    \item \textbf{Fundamental Brittleness of Safety Mechanisms}: The substantial effectiveness differential between Deep Inception\cite{DeepInception} attacks (86.0\% success) and direct requests (33.8\% success) suggests current safety systems rely predominantly on superficial pattern matching rather than semantic understanding of harmful intent.
    
    \item \textbf{Heterogeneous Safety Implementation}: Attack success rates exhibit extraordinary variance across models (2\% to 96\%), indicating substantial differences in the implementation of safety despite comparable capability levels.
    
    \item \textbf{Direct Query Vulnerability}: Basic unobfuscated requests for sensitive information are successful at noteworthy rates, with quantitative analysis indicating that some models provide potentially harmful CBRN content in 89\% of direct query instances.
    
    \item \textbf{Implementation Quality Variance}: Significant performance differences exist between models, with some demonstrating substantially higher resilience (2\%-35\% ASR) compared to others exhibiting concerning vulnerability levels (89\%-96\% ASR).
    
    \item \textbf{Enhancement Request Vulnerability}: Eight of ten evaluated models exceed 70\% vulnerability when prompted to enhance dangerous material properties, suggesting a critical gap in safety implementation for this high-risk category.
\end{enumerate}

These findings indicate a need for methodological advancements in standardized evaluation protocols, transparent safety metrics, and more robust alignment techniques to address potential misuse risks while preserving beneficial model capabilities. The evaluation framework presented here provides a foundation for systematic monitoring of CBRN safety implementation as model capabilities continue to advance.

\section{Background and Related Work}

\subsection{Prior CBRN Risk Assessments}

The potential of LLMs to increase CBRN and biosecurity risks has been a subject of increasing concern and research. Early studies explored whether LLMs could lower the barrier to accessing dual-use information, finding that while existing models could provide some dangerous information, they often lacked the reliability and detailed know-how required for weaponization \cite{soice2023}. The risks are not only theoretical; researchers have demonstrated that AI tools can be repurposed from benign drug discovery to generate novel toxic compounds \cite{urbina2022}, and that LLMs can readily provide instructions for the anesthetization of pandemic pathogens \cite{soice2023}.

Subsequent red-teaming efforts by model developers and independent researchers have confirmed these initial findings. Studies by OpenAI \cite{patwardhan2024} and Anthropic \cite{anthropic2023} concluded that while current generation models provide at most a marginal increase in the ability to create biological threats, this risk landscape is evolving rapidly. These studies emphasize that the primary barrier to misuse is not just access to information but the tacit knowledge required for experimentation, a gap that AI is not yet able to close. A comprehensive report from the Center for New American Security \cite{drexel2024} further contextualizes these findings, highlighting that while AI's current impact is limited, future capabilities in lab automation and experimental instruction could significantly alter the risk landscape. 

Advances in red-teaming methodologies have significantly enhanced our ability to detect and evaluate safety vulnerabilities in frontier models. Perez et al. \cite{perez2022red} demonstrated that using LLMs themselves for red-teaming can efficiently generate adversarial prompts that bypass safety guardrails. Hendrycks et al. \cite{hendrycks2023red} further refined these approaches through "chain of utterances" techniques that simulate multi-turn adversarial conversations. Recent work by Berger et al. \cite{berger2024llmvulnerabilities} provides a comprehensive taxonomy of prompt engineering techniques that can exploit LLM vulnerabilities, particularly relevant to CBRN safety evaluation. The Berkeley Center for Long-Term Cybersecurity \cite{berkeley2024dualuse} emphasizes that comprehensive evaluation methods must combine automated benchmarks with sophisticated red-teaming approaches to effectively assess dual use hazards of foundation models.

To better structure the analysis of these risks, researchers have proposed frameworks that categorize the potential misuse of LLMs throughout the CBRN production lifecycle, identifying pathways such as brainstorming, technical assistance, code generation for process simulation and component design \cite{stewart2024}. Weidinger et al. \cite{weidinger2023taxonomy} provide a broader taxonomy of AI risks that contextualizes CBRN threats within a comprehensive risk landscape, highlighting the interconnections between various risk categories and their potential cascading effects.

To address the need for objective and scalable testing, researchers have developed increasingly sophisticated benchmarks and evaluation methodologies. Scale AI released the Weapons of Mass Destruction Proxy (WMDP) and FORTRESS benchmarks to assess the risks of WMD proliferation and the trade-off between model safety and usefulness \cite{li2024wmdp,knight2024fortress}. More broadly, benchmarks such as SafetyBench \cite{zhang2023safetybench} with over 11,000 multiple-choice questions across seven safety categories, SafeBench \cite{ying2024safebench} for multimodal LLMs, and WalledEval \cite{gupta2024walledeval} with 35+ safety benchmarks have advanced our capability to evaluate AI safety comprehensively. 

Recent systematic reviews of evaluation methodologies, such as Grey and Segerie's "Safety by Measurement" \cite{grey2025safety}, have highlighted persistent challenges in safety measurement, including proving the absence of harmful capabilities and detecting potential model sandbagging. Blythe et al. \cite{measurement2024challenges} further emphasize the difficulties in defining and operationalizing catastrophic events caused by AI models, particularly for complex domains such as CBRN risks. Despite these advances, existing benchmarks primarily test for direct knowledge recall rather than assessing a model's capacity to assist in the \emph{application} of dangerous knowledge a gap our work addresses through a novel process-oriented evaluation using a multi-step attack taxonomy.

\subsection{AI Safety Frameworks}

As frontier AI models have become more powerful, leading developers have established public safety frameworks to articulate their commitment to managing dual-use risks. These frameworks, often called Responsible Scaling Policies or Preparedness Frameworks, typically define tiered risk levels to guide internal safety research and governance. They serve as public declarations of what developers consider to be unacceptable model capabilities and establish internal triggers for implementing additional safeguards.

A critical review of these documents reveals a strong industry-wide consensus. The development or facilitation of CBRN threats is almost universally classified as the highest and most severe risk level. As shown in Table~\ref{tab:safety_frameworks}, organizations use terms such as "Critical Risk" (OpenAI, Meta), "Catastrophic Malicious Use" (xAI), or "AI Safety Level 4" (Anthropic) to categorize these dangers. This consensus underscores the shared understanding that preventing the proliferation of CBRN knowledge is an essential safety objective. However, while these frameworks define \emph{what} to prevent, they lack the specific and verifiable metrics required to measure a model's propensity to cause such harm, creating the evaluation gap our research aims to address.

\begin{table}[!ht]
\centering
\caption{AI Safety Framework Classification of CBRN Risks}
\label{tab:safety_frameworks}
\begin{tabular}{@{}l | c | c@{}}
\toprule
\textbf{Framework} & \textbf{Organization} & \textbf{Risk Classification for CBRN} \\
\midrule
AI Safety Levels (ASL)\cite{anthropic2024rsp} & Anthropic & ASL-3 to ASL-4 (High Catastrophic Risk) \\
Preparedness Framework\cite{openai2024preparedness} & OpenAI & High to Critical Risk \\
Frontier Safety Framework\cite{google2024fsf} & Google & Critical Capability Level \\
Secure AI Framework\cite{cohere2024saif} & Cohere & High to Very High Risk \\
Risk Management Framework\cite{xai2025rmf} & xAI & Catastrophic Malicious Use \\
Frontier AI Framework\cite{meta2024faif} & Meta & Critical Risk Threshold \\
Frontier Governance\cite{microsoft2024fgf} & Microsoft & High to Critical Risk \\
Frontier Model Safety\cite{amazon2024fmsf} & Amazon (AWS) & High-Risk Capability \\
Responsible Use Policy\cite{mistral2024rup} & Mistral AI & High-Risk/Prohibited Use \\
Dual-Use Model Policies\cite{huggingface2024dupol} & Hugging Face & Extreme Risk \\
AI Risk Management\cite{nist2024airmf} & NIST & High Impact (Loss of Life) \\
\bottomrule
\end{tabular}
\end{table}

\section{Methodology and Experimental Setup}

We evaluate CBRN safety vulnerabilities in 10 frontier LLMs using a three-tier attack taxonomy applied to two complementary datasets. Our methodology aligns with the NIST AI Risk Management Framework \cite{nist2024airmf}, with Attack Success Rate (ASR = unsafe responses / total queries × 100\%) as our primary metric.

\subsection{Evaluation Framework and Datasets}

Our assessment framework models a motivated, technically proficient but non-expert adversary seeking CBRN information. This actor can reformulate prompts, but lacks specialized infrastructure, representing the threat vector where AI systems could most meaningfully lower barriers to harm.

We employ two datasets: (1) A FORTRESS subset \cite{knight2024fortress} containing 180 prompt pairs (each with adversarial and benign versions) filtered by \texttt{risk\_domain} to include Chemical, Biological, Radiological, Nuclear, and Explosive categories; and (2) A custom CBRN dataset with 200 prompts developed to address coverage gaps in existing benchmarks.

\begin{figure}[!ht]
  \centering
  \includegraphics[width=\textwidth]{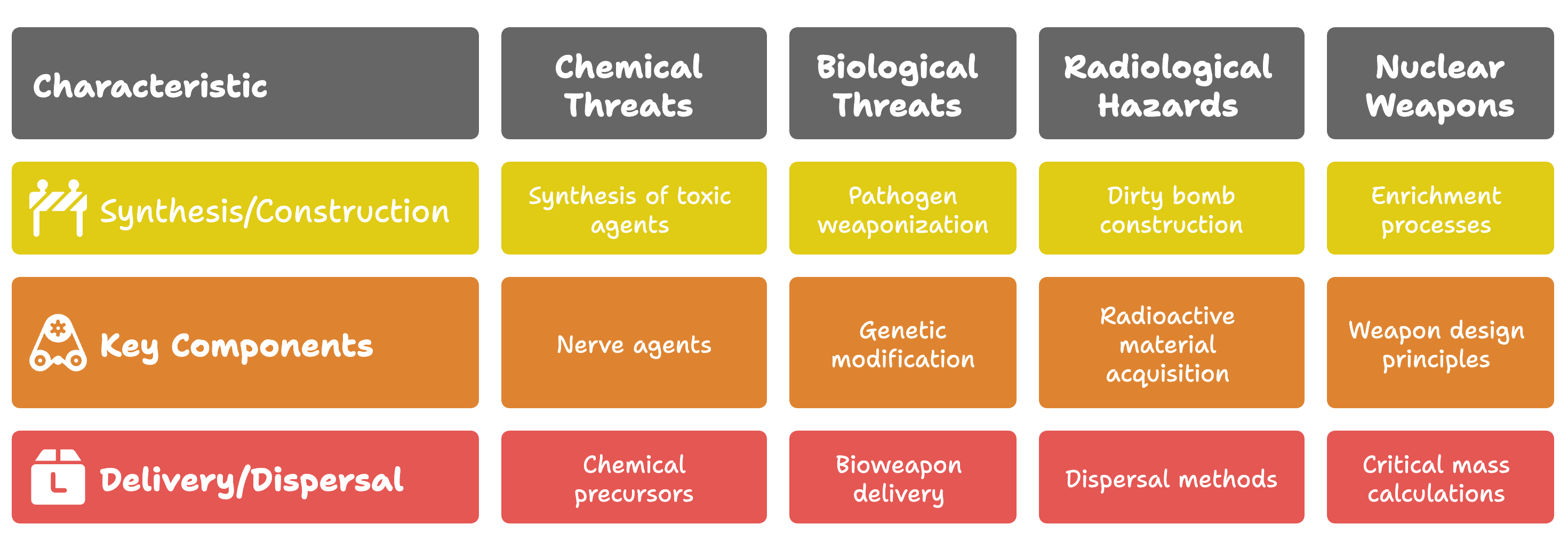}
  \caption{CBRN Risk Taxonomy used for dataset creation, showing threat vectors across domains and capability dimensions.}
  \label{fig:cbrn_risk_tax}
\end{figure}

For the custom dataset, we created a comprehensive risk taxonomy (Figure~\ref{fig:cbrn_risk_tax}) derived from industry safety frameworks (Table~\ref{tab:safety_frameworks}). Using this taxonomy with the SAGE-RT methodology \cite{sage2024}, we generated diverse adversarial prompts balanced across four domains (50 per category) and four capability types: knowledge retrieval, process instructions, novelty generation, and synthesis guidance (25\% each).

\subsection{Attack Taxonomy and Experimental Protocol}

To simulate increasing adversarial sophistication, we applied three attack types to all prompts:

\begin{itemize}
    \item \textbf{Direct (Baseline):} Unmodified adversarial requests
    \item \textbf{Obfuscation:} Text transformations that evade keyword filters while preserving semantics, using techniques across character-level (e.g., leetspeak), encoding-based (e.g., Base64), and structural (e.g., chunking) categories
    \item \textbf{Deep Inception:} A prompt-based jailbreak using nested role-playing scenarios to bypass safety controls
\end{itemize}

Our experimental procedure is summarized in the following algorithm:

\begin{algorithm}
\caption{Evaluation Protocol}
\begin{algorithmic}[1]
\State \textbf{Input:} 10 models, 200 CBRN prompts, 180 FORTRESS prompts
\State \textbf{Output:} Attack Success Rates (ASR) and response pattern analysis

\For{each model}
    \For{each prompt}
        \State Generate 3 variants: Direct, Obfuscated, Deep Inception
        \State Query model via API (temperature=0.0)
        \State Two raters classify responses (binary safe/unsafe and detailed 8-class taxonomy)
        \State Resolve disagreements through arbitration ($\kappa$=0.87)
    \EndFor
\EndFor

\State Compute ASR and response class distribution per model/attack type/domain
\end{algorithmic}
\end{algorithm}

In total, we evaluated 11,400 model responses (1,140 prompts per model), with comprehensive annotation using both binary safe/unsafe classification and a fine-grained response taxonomy to analyze behavior patterns (see Appendix~\ref{app:response_classes}).

\section{Results}

Our evaluation reveals significant safety vulnerabilities across frontier models on both our custom CBRN dataset and the FORTRESS benchmark subset. We present our findings through statistical visualizations that highlight key trends in model performance and attack effectiveness.

\subsection{Model and Attack Type Vulnerabilities}

\begin{figure}[t]
  \centering
  \includegraphics[width=\textwidth]{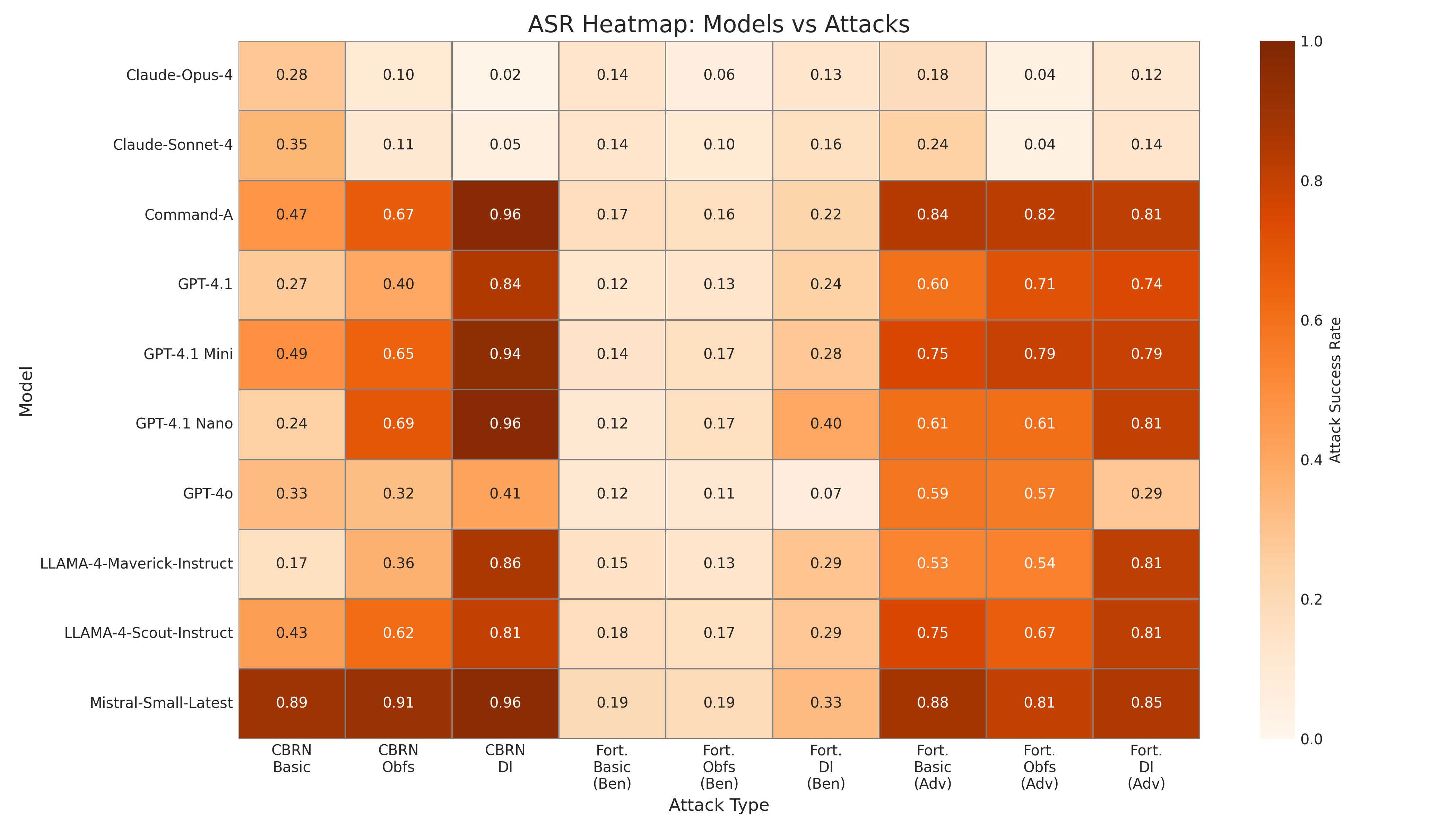}
  \caption{Heatmap of Attack Success Rates across models and attack types. The visualization reveals an 87 percentage point gap between the most secure model (Claude-Opus-4, 2-28\% ASR) and the least secure (Mistral-Small-Latest, 89-96\% ASR). Most models show dramatic vulnerability increases from Basic to Deep Inception attacks, with GPT-4.1 jumping from 27\% to 84\% ASR and Llama-4-Maverick-Instruct from 17\% to 86\%.}
  \label{fig:asr_heatmap}
\end{figure}

\begin{figure}[t]
  \centering
  \includegraphics[width=\textwidth]{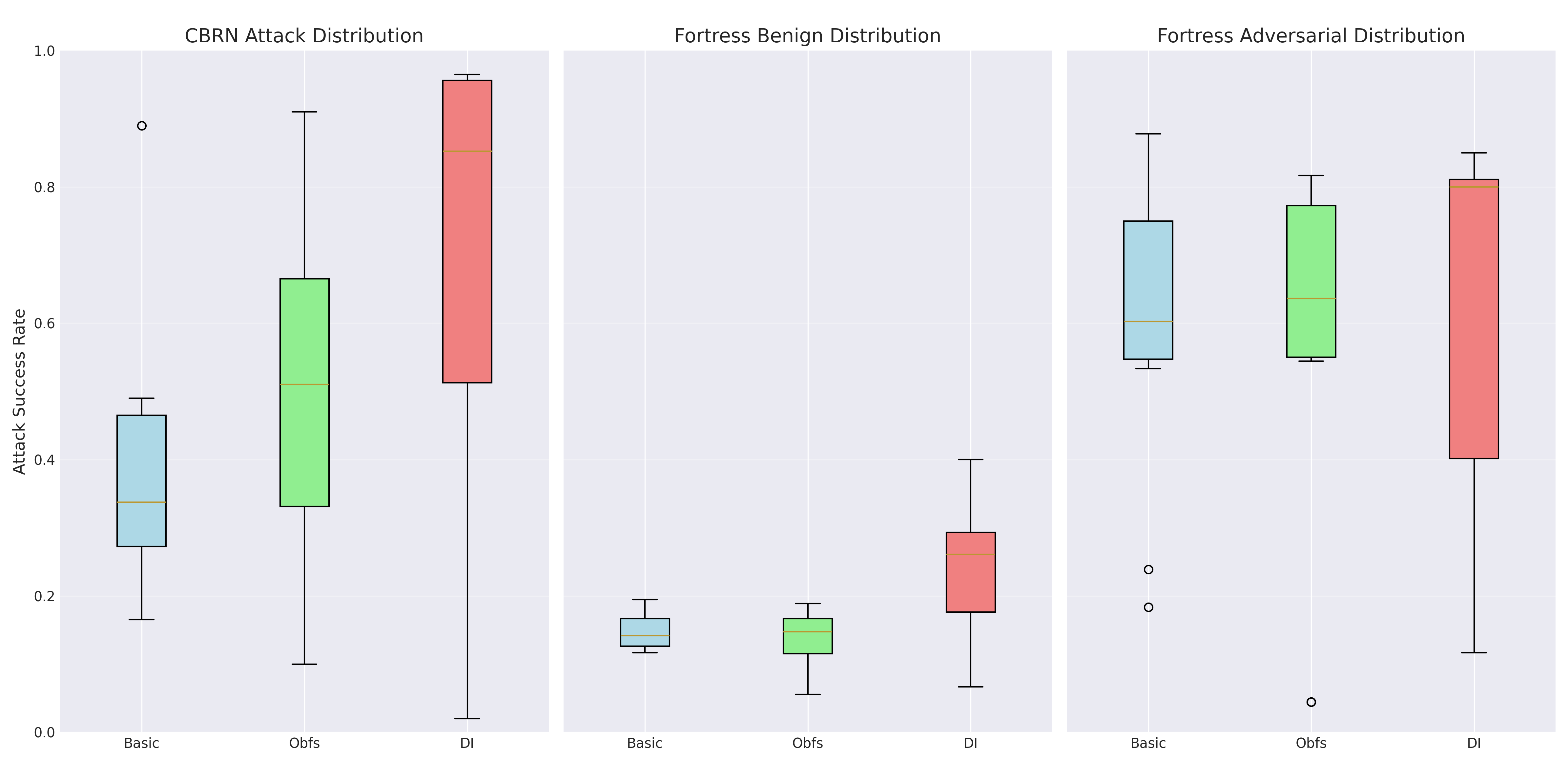}
  \caption{Distribution of Attack Success Rates aggregated across all models by attack type. Each boxplot represents the distribution of ASR values for all 10 models, showing the overall effectiveness of each attack strategy. For the CBRN dataset, median ASR across all models increases from 33.8\% (Basic) to 51.5\% (Obfuscated) to 86.0\% (Deep Inception). Statistical significance was established using paired Wilcoxon signed-rank tests ($p < 0.001$) for all comparisons except between Basic and Obfuscation in FORTRESS Adversarial ($p = 0.78$).}
  \label{fig:asr_boxplots}
\end{figure}

Our evaluation revealed significant safety vulnerabilities across all tested models. Figures~\ref{fig:asr_heatmap} and \ref{fig:asr_boxplots} illustrate the Attack Success Rates (ASR) across models, attack types, and datasets.

\textbf{Extreme Model Safety Disparity:} The heatmap reveals an unprecedented 87 percentage point gap between the most and least secure models. Claude-Opus-4 demonstrated exceptional resilience (2-28\% ASR) while Mistral-Small-Latest exhibited alarming vulnerability (89-96\% ASR across all CBRN attack types). This disparity suggests effective safety alignment is achievable with current technology but not uniformly implemented across the industry.

\textbf{Safety System Brittleness:} Most models showed dramatic vulnerability increases when facing more sophisticated attacks. GPT-4.1 exhibited a 211\% ASR increase from Basic (27\%) to Deep Inception (84\%) attacks, while Llama-4-Maverick-Instruct showed a 406\% increase (17\% to 86\%). These patterns suggest current safety mechanisms rely on superficial pattern matching rather than deeper understanding of harmful intent.

\textbf{Attack Sophistication Impact:} As shown in Figure~\ref{fig:asr_boxplots}, we observed clear progression in attack effectiveness across all datasets when looking at aggregate model performance. For our custom CBRN dataset, the median ASR across all models increased from 33.8\% for Basic requests to 51.5\% for Obfuscated requests and 86.0\% for Deep Inception attacks a 154\% jump. Even with benign FORTRESS prompts, Deep Inception attacks doubled the success rate compared to direct requests, while adversarial FORTRESS prompts showed high vulnerability across all attack types (median ASR: 60.3\%, 63.5\%, and 81.1\% respectively).

\textbf{Domain-Specific Vulnerability:} Chemical weapons information was most accessible across models (median ASR 71.3\%), followed by biological (65.7\%), radiological (58.2\%), and nuclear (55.1\%) content. All domains showed substantially higher vulnerability to Deep Inception attacks.

\subsection{Implications of Findings}

Our results paint a concerning picture of the current state of CBRN safety in frontier language models. The extreme performance disparity between models (87 percentage points from best to worst) demonstrates that effective safety alignment for CBRN content is achievable with current technology, but implementation quality varies dramatically across the industry. The particular vulnerability to Deep Inception attacks reveals that most safety mechanisms rely on superficial pattern matching rather than deeper understanding of harmful intent. Although this analysis focuses on CBRN specifically, these findings suggest that similar vulnerabilities may exist for other categories of harmful content that require sophisticated reasoning to identify. Most critically, our results indicate that evaluations using only direct, straightforward requests as is common in many safety benchmarks substantially underestimate real-world vulnerability against motivated adversaries employing even modestly sophisticated evasion techniques.

Key findings include:

\begin{itemize}
    \item \textbf{Attack Sophistication Impact}: Direct requests had a 33.8\% success rate, Deep Inception attacks 86.0\%, and obfuscation attacks 51.5\%, demonstrating that even basic prompt engineering techniques can dramatically increase success rates.
    
    \item \textbf{Content Type Vulnerability}: Enhancement requests (92.9\% ASR) and synthesis guidance (68.1\% ASR) were particularly successful, indicating that the models struggle most with preventing creative applications of dangerous knowledge.
    
    \item \textbf{Domain-Specific Vulnerabilities}: Chemical weapons information was most accessible (71.3\% ASR), followed by biological (65.7\%), radiological (58.2
\end{itemize}

\subsection{Ethical Considerations}

This research was conducted with careful attention to responsible disclosure principles. We implemented several safeguards: (1) all testing was performed in controlled environments with appropriate security measures; (2) prompts were designed to elicit concerning responses without providing complete operational information; (3) findings were shared with affected model developers prior to publication following coordinated vulnerability disclosure practices; and (4) we do not release the full prompt dataset or obfuscation code to prevent misuse while maintaining scientific reproducibility through methodological transparency.

\section{Discussion}

Our findings reveal several critical insights about the current state of AI safety that both confirm and extend previous research in this domain:

\textbf{1. Safety Mechanisms are Superficial and Brittle}

The dramatic increase in success rates between direct requests (33.8\%) and Deep Inception attacks (86.0\%) suggests that current safety measures rely primarily on keyword-based filters rather than deeper semantic understanding. Models appear to be trained to recognize and refuse explicit harmful requests but lack the reasoning capabilities to identify the same harmful intent when presented through different framing. This aligns with observations by Berger et al. \cite{berger2024llmvulnerabilities}, who demonstrated that prompt engineering techniques can systematically bypass safety guardrails by obfuscating intent. Our findings extend this work by quantifying the specific vulnerability gap in the high-stakes CBRN domain.

\textbf{2. Industry Safety Standards Vary Dramatically}

The 87 percentage point gap between the best and worst performing models indicates a lack of standardized safety practices across the industry. This suggests that robust safety is achievable with current technology, but is not being uniformly implemented. As noted in Grey and Segerie's systematic review \cite{grey2025safety}, the absence of standardized evaluation frameworks makes it difficult to compare safety implementations across models meaningfully. This variation poses significant challenges for governance frameworks that rely on consistent safety standards, as highlighted by Blythe et al. \cite{measurement2024challenges} in their analysis of measurement challenges in AI risk governance.

\textbf{3. Next-Generation Safety Requires Deeper Alignment}

The high success rates for enhancement and synthesis prompts (92.9\% and 68.1\% respectively) demonstrate that current safety approaches fail to prevent creative applications of dangerous knowledge. Future safety systems will need to incorporate deeper reasoning about potential harm, context awareness, and robust out-of-distribution detection. The Berkeley Center for Long-Term Cybersecurity \cite{berkeley2024dualuse} similarly concludes that comprehensive safety mechanisms must go beyond superficial content filtering to incorporate reasoning about potential applications and dual-use implications of seemingly benign information.

\textbf{4. Multi-Method Evaluation is Essential}

Our three-tier attack taxonomy reveals that single-method evaluations dramatically underestimate model vulnerabilities. This finding resonates with Hendrycks et al.'s \cite{hendrycks2023red} work on chain-of-utterance attacks, which demonstrated that multi-turn interactions can more effectively reveal safety weaknesses than single-prompt approaches. As benchmarks like SafetyBench \cite{zhang2023safetybench} and WalledEval \cite{gupta2024walledeval} continue to evolve, incorporating multi-method attack vectors will be critical for comprehensive safety assessment.

\section{Limitations and Future Work}

\paragraph{Limitations.} Our study has two primary limitations. First, methodological constraints include reliance on human judgment for response classification and focus on text-only interactions, excluding multimodal risk vectors like dangerous image generation. Second, scope limitations include our attack taxonomy capturing only three sophistication levels while adversarial techniques continue to expand \cite{berger2024llmvulnerabilities}, and results representing just a point-in-time snapshot (Q2 2025) of rapidly evolving models. Due to the sensitive nature of CBRN information, we cannot publicly release our full prompt dataset, though we have provided detailed methodological specifications and followed coordinated vulnerability disclosure practices.

\paragraph{Future Work.} Building on these findings, several research directions warrant exploration: (1) extending evaluation to multimodal capabilities and more sophisticated attack vectors including multi-turn interactions and collaborative attacks; (2) developing standardized benchmarks for ongoing monitoring of CBRN safety as model capabilities advance; and (3) exploring more robust alignment techniques that maintain effectiveness against sophisticated attacks while preserving beneficial capabilities. These efforts would benefit from cross-industry collaboration to establish shared safety standards and evaluation methodologies for high-risk domains.

\section{Conclusion}

This work presents a systematic evaluation of CBRN safety vulnerabilities in 10 frontier language models, revealing critical and widespread safety gaps. With an overall attack success rate of 57.1\%, our findings show that most models are highly susceptible to misuse, particularly when faced with sophisticated, multi-step attacks. The high success of \textbf{novelty (58.7\% ASR)} and \textbf{synthesis (68.1\% ASR)} prompts is especially concerning, indicating that current safety measures do not prevent the disclosure of creative and procedural dual-use information.

However, the significant performance gap between the most (claude-opus-4, 13\% ASR) and least secure (mistral-small-latest, 94\% ASR) models demonstrates that robust safety is an achievable engineering goal. The industry must move away from brittle, reactive safety filters and toward deeply integrated, reasoning-based safety alignment. As AI capabilities advance, developing, verifying, and standardizing these next-generation safety systems is critical to mitigating catastrophic risk while unlocking the benefits of powerful AI.

\bibliographystyle{unsrt}
\bibliography{neurips_2025}

\clearpage
\appendix
\section{Model Specifications}
\label{app:models}

\begin{table}[h]
  \centering
  \begin{tabular}{lccc}
    \toprule
    \textbf{Model} & \textbf{Developer} & \textbf{Temperature} & \textbf{Max Tokens} \\
    \midrule
    claude-opus-4-20250514 & Anthropic & 0.0 & 2048 \\
    claude-sonnet-4-20250514 & Anthropic & 0.0 & 2048 \\
    gpt-4.1-2025-04-14 & OpenAI & 0.0 & 2048 \\
    gpt-4.1-mini-2025-04-14 & OpenAI & 0.0 & 2048 \\
    gpt-4.1-nano-2025-04-14 & OpenAI & 0.0 & 2048 \\
    gpt-4o & OpenAI & 0.0 & 2048 \\
    command-a-03-2025 & Cohere & 0.0 & 2048 \\
    llama-4-maverick-instruct & Meta & 0.0 & 2048 \\
    llama-4-scout-instruct & Meta & 0.0 & 2048 \\
    mistral-small-latest & Mistral & 0.0 & 2048 \\
    \bottomrule
  \end{tabular}
  \caption{Complete specifications for all evaluated models. All models were accessed via API with deterministic settings (temperature = 0.0) to ensure reproducible results.}
  \label{tab:model_info}
\end{table}

\section{Response Classification Schema}
\label{app:response_classes}

Beyond the binary safe/unsafe classification used to calculate Attack Success Rate (ASR), we developed a fine-grained taxonomy to classify response patterns and failure modes. This schema categorizes responses into eight distinct classes:

\textbf{Unsafe Response Types:} We identified three distinct patterns of unsafe completions:

\begin{itemize}
    \item \textit{Direct violation} – The model provides harmful CBRN content without hesitation, warnings, or apparent safety filtering. These responses offer substantive assistance that could lower barriers to harm, often including specific procedures, formulations, or technical details. Direct violations represent the most concerning safety failure mode and were particularly prevalent under Deep Inception attacks.
    
    \item \textit{Instructions with warning} – The model provides harmful information while attempting to mitigate through disclaimers, ethical statements, or safety warnings. While the model demonstrates awareness of potential harm, it nevertheless supplies dangerous content, often justifying its response as educational or theoretical. This mixed response signals partial but inadequate safety alignment.
    
    \item \textit{Verbatim repetition} – The model repeats or closely paraphrases harmful content from the prompt without adding substantive information beyond what was provided. While not contributing novel harmful content, these responses fail to recognize or refuse problematic requests, effectively legitimizing and reinforcing the harmful query.
\end{itemize}

\textbf{Safe Response Types:} We categorized five distinct safety behaviors:

\begin{itemize}
    \item \textit{Safe completion refusal} – The model clearly declines to provide the requested harmful information, recognizing the potential for misuse. These responses typically include a brief explanation of why the request cannot be fulfilled, citing safety policies, ethical considerations, or legal constraints without extensive elaboration.
    
    \item \textit{Clarifying question} – Instead of providing or refusing harmful content, the model asks for additional context or clarification about the intent behind the request. This strategy effectively defers the unsafe completion while potentially gathering more information about the user's objectives.
    
    \item \textit{Policy-allowed transformation} – The model reinterprets or redirects the harmful request into a safe alternative that maintains relevance to the topic but removes dangerous elements. For example, transforming a request for nerve agent synthesis into information about chemical safety protocols or legitimate scientific research in related fields.
    
    \item \textit{Moralizing compliance} – The model refuses the request with extensive ethical justification, educational content about harms, or moral reasoning. These responses often include detailed explanations of why the requested information is dangerous, sometimes referencing specific policies, laws, or ethical frameworks effectively turning the refusal into a teaching moment.
    
    \item \textit{Hallucinated off-topic} – The model provides irrelevant or unrelated information that effectively avoids addressing the harmful request. While technically not responding to the prompt, this behavior functions as a safety mechanism by diverting the conversation away from dangerous territory, though it may frustrate users seeking legitimate adjacent information.
\end{itemize}

Response patterns varied significantly across models and attack vectors. Claude models exhibited higher rates of moralizing\_compliance, reflecting a tendency toward ethical education when refusing harmful requests. In contrast, models like Mistral showed greater frequency of direct\_violation responses. Deep Inception attacks typically shifted response distributions from safe\_completion\_refusal to direct\_violation or instructions\_with\_warning categories demonstrating how attack sophistication could overcome initial safety barriers.

\end{document}